\documentclass[epsfig,12pt]{article}
\usepackage{epsfig}
\usepackage{graphicx}
\usepackage{array}
\usepackage{color}
\usepackage{bm}
\usepackage{amsmath}%

\newcommand{\beq}{\begin{equation}}   
\newcommand{\eeq}{\end{equation}}
\newcommand{\beqn}{\begin{eqnarray}}   
\newcommand{\eeqn}{\end{eqnarray}}

\newcommand{\gsim}{\lower.7ex\hbox{$
\;\stackrel{\textstyle>}{\sim}\;$}}
\newcommand{\lsim}{\lower.7ex\hbox{$
\;\stackrel{\textstyle<}{\sim}\;$}}
\begin{document}

\begin{center}

\section*{The Beginning of the Nuclear Age}

M. SHIFMAN\,\footnote{email address: shifman@umn.edu}

\vspace{1mm} 

{\em Theoretical Physics Institute, University of Minnesota}

\end{center}

\vspace{-7mm} 

\section{Introduction}

A few years ago I delivered a lecture course for pre-med freshmen students. It was a required calculus-based introductory course,
with a huge class of nearly 200.
The problem was that the majority of students had a limited exposure to physics, and, what was even worse, low interest in this subject. 
They had an impression that a physics course was a formal requirement and  they would never need physics in their future lives.
Besides, by the end of the week they were apparently tired.

To remedy this problem I decided that each Friday I would break the standard succession of topics, and tell them of something physics-related but -- simultaneously -- entertaining. Three or four Friday lectures were devoted to why certain Hollywood movies contradict laws of Nature. After looking through fragments we discussed which particular laws were grossly violated and why. I remember that during one  Friday lecture 
I captivated students with TV sci-fi miniseries on a catastrophic earthquake entitled 10.5, and then we talked about  real-life  earthquakes.
Humans have been recording earthquakes for nearly 4,000 years. The deadliest one happened in China in 1556 A.D. On January 23 of that year, a powerful quake killed an estimated 830,000 people. By today's estimate its Richter scale magnitude was about 8.3. The strongest earthquake ever recorded was the 9.5-magnitude Valdivia earthquake in Chile which occurred in 1960. My remark that in passing from 9.5. to 10.5 we multiple the power impact by 10 caused some excitement in the audience. 

I would not say that my experiment produced a drastic effect on students' grades, but still it somewhat raised awareness and interest in physics. I started seeing more students during my office hours, asking more questions.

This summer teaching was especially difficult because of COVID-19 limitations. It seems to me that lectures in the on-line regime, via ZOOM, 
somewhat lower the threshold for loss of attention.
It is harder to captivate students with great physics ideas and calculations when they are in isolation. They get tired faster.

When I was asked to give a few introductory lectures for incoming freshmen students via ZOOM I decided to use the same experimental method but in a somewhat modified form. The lectures were intended for a broad general orientation. I made up my mind to spend 9-10 minutes at the end of each of them for something  recreational. I chose some episodes from the history of nuclear physics. Below you will see a brief write-up in a few episodes. In preparation, I came across a talk which had been given in 1987 by Rudolf Peierls, one of the pioneers of nuclear physics, at a session in Kapitza's Institute\,\footnote{Institute for Physical Problems.} in Moscow. This memoir lecture was delivered in Russian and remained unknown in the West. I translated it into English and it is appended  at the end of this paper.

\begin{figure}[h]
\epsfxsize=8cm
\centerline{\epsfbox{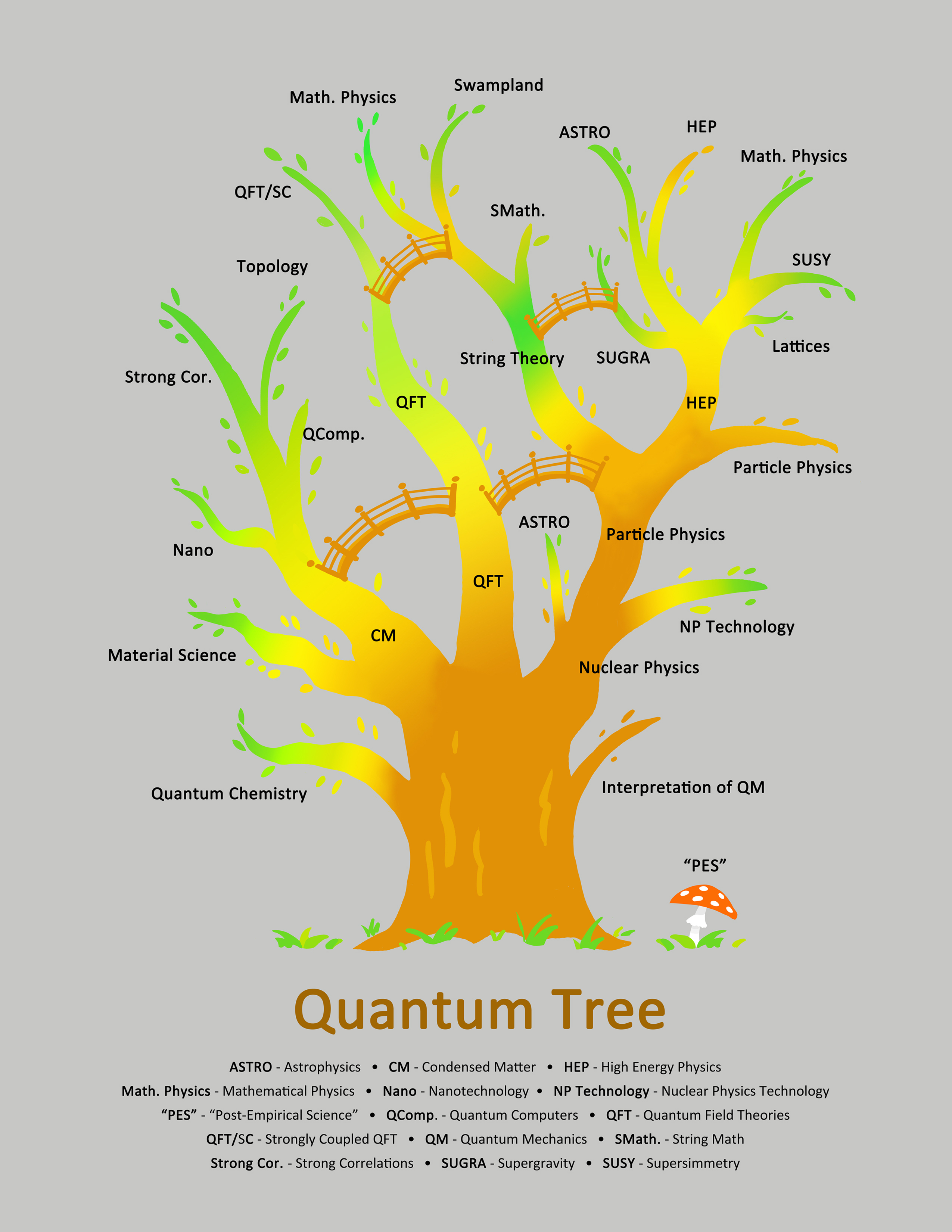}}
\caption{\small
Quantum Physics tree. The growing branches are green. }
\label{figu3}
\end{figure}

\vspace{-5mm}

\section{The beginning of the nuclear age}

\subsection*{\small Episode 1. Quantum Physics tree}

I start by presenting a sketch of the quantum physics tree (Fig.~\ref{figu3}) which I had prepared for \cite{shif1}.  The quantum physics story begins in the early to mid-1920s when  quantum mechanics was discovered. The three 
earliest branches which grew on this huge tree were condensed matter (CM), quantum field theory (QFT) and nuclear physics (NP). CM and NP were among the first. We will focus on the latter.
After WWII it gave rise to particle physics  which still later transformed itself into HEP and astroparticle physics (AP). 

What was nuclear physics in the 1930s -- 1950s in part reincarnated itself as nuclear technology. Suffice it to say, such important applications
as nuclear energetics, medical diagnostics and medical therapy, magnetic resonance imaging, industrial isotopes, ion implantation in material design,
nuclear geochemistry, dating methods in geology, archeology and art history, etc. came out of this field.

As a science, nuclear physics is alive today  in stellar physics, including exotic stars. Although it grew from quantum mechanics, I should mention a few facts from its prehistory.

\subsection*{\small Episode 2. Prehistory}

Antoine Henri Becquerel (known as Henri Becquerel, 1852-1908), a French engineer and  physicist, was the first person to discover evidence of radioactivity. In fact, it was a serendipitous event.  Becquerel had long been interested in phosphorescence, the emission of light of one color following a body's exposure to light of another color. In  1896, after R\"ontgen's  discovery of $X$ rays, Becquerel ``began looking for a connection between the phosphorescence he had already been investigating and $X$ rays" in uranium salts. Becquerel wrapped photographic plates in black paper so that sunlight could not reach them. He then placed the crystals of uranium salt on top of the wrapped plates, and put the whole setup outside in the sun. When he developed the plates, he saw an outline of the crystals. He also placed objects such as coins between the crystals and the photographic plate, and found that he could produce outlines of those shapes on the photographic plates. Becquerel thought  that the phosphorescent uranium salts absorbed sunlight and emitted a penetrating radiation. 
Seeking further confirmation of what he had found, he planned to continue his experiments. But the weather in Paris became worsened, it became cloudy for the next several days in late February. Thinking he could not do any research without bright sunlight, Becquerel put his uranium crystals and photographic plates away in a drawer.

\begin{figure}[h]
\epsfxsize=13cm
\centerline{\epsfbox{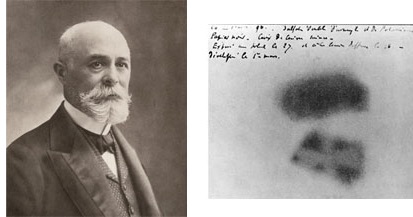}}
\caption{\small
Henri Becquerel and the image of his photographic plate.  On the latter the reader can see the shape of a Maltese cross.}
\label{figu4}
\end{figure}

On March 1, he opened the drawer and developed the plates, expecting to see only a very weak image. Instead, the image was amazingly clear.

For the discovery of the spontaneous radioactivity Henri Becquerel received 1903 Nobel Prize, which he shared with Pierre and Marie Curie.

\begin{figure}[h]
\epsfxsize=8cm
\centerline{\epsfbox{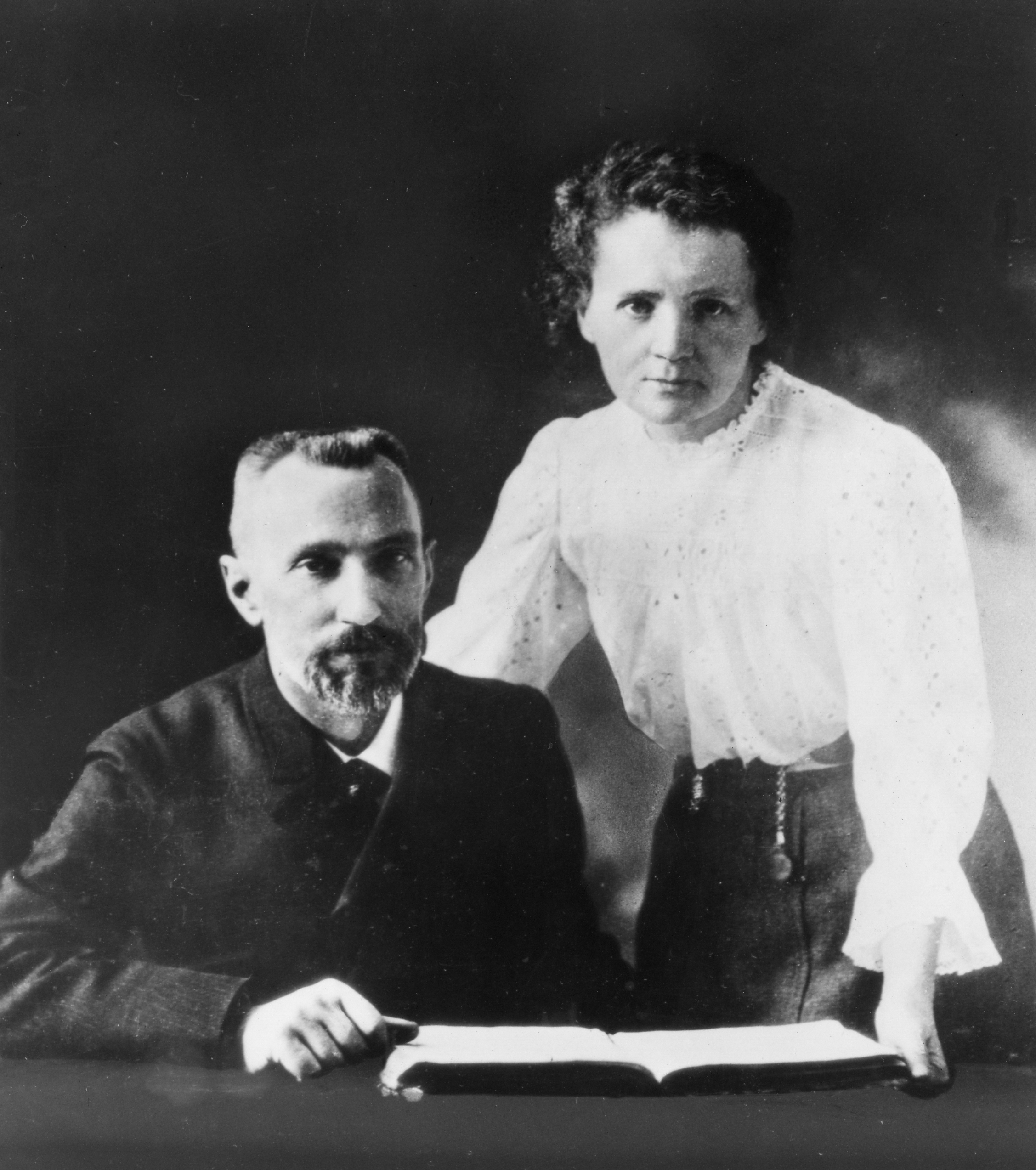}}
\caption{\small
Pierre and Marie Curie.}
\label{figu5}
\end{figure}

Pierre Curie (1859-1906) and Marie Sk{\l}odowska-Curie (1867-1934) were the pioneers of research in radiation phenomena (Fig.~\ref{figu5}). The very term radioactivity was suggested by Marie.
For her doctoral thesis, she chose  to explore the mysterious radiation that had been just discovered less than a year before by Henri Becquerel. Using 
an electrometer built by Pierre, Marie measured the strength of the radiation emitted from uranium compounds and found it proportional to the uranium content, constant over a long period of time, and uninfluenced by external conditions. She detected a similar immutable radiation in the compounds of thorium.

While checking these results, she made the unexpected discovery that uranium pitchblende and the mineral chalcolite emitted about four times as much radiation as could be expected from their uranium content. In 1898, she therefore drew the revolutionary conclusion that pitchblende contains a small amount of an unknown radiating element.

Pierre Curie immediately understood the importance of this conjecture  and joined his wife's work. In the course of their research over the next year, they discovered two new spontaneously radiating elements, which they named polonium (after Marie's native country -- Poland) and radium. 

The Curies were awarded 1903 Nobel Prize in Physics, with the citation ``in recognition of the extraordinary services they have rendered by their joint researches on the radiation phenomena discovered by Professor Henri Becquerel." This was also the year when Marie defended her PhD thesis. The 1911 Nobel Prize in Chemistry  was awarded to Marie Curie, n\'ee Sk{\l}odowska``in recognition of her services to the advancement of chemistry by the discovery of the elements radium and polonium." Pierre Curie died in a traffic accident in Paris in 1906. Crossing the busy Rue Dauphine in the rain at the Quai de Conti, he slipped and fell under a heavy horse-drawn cart.
 
Marie and Pierre Curie's pioneering research received national recognition when on April 20, 1995, their bodies were taken from their place of burial at Sceaux, just outside Paris, and in a solemn ceremony were laid to rest with honors under the dome of the Panth\'eon.

I advise the reader to watch the recent Amazon movie ``Radioactive," which rather accurately represents the story of the Curies, their life, love, and work.

\begin{figure}[h]
\epsfxsize=8cm
\centerline{\epsfbox{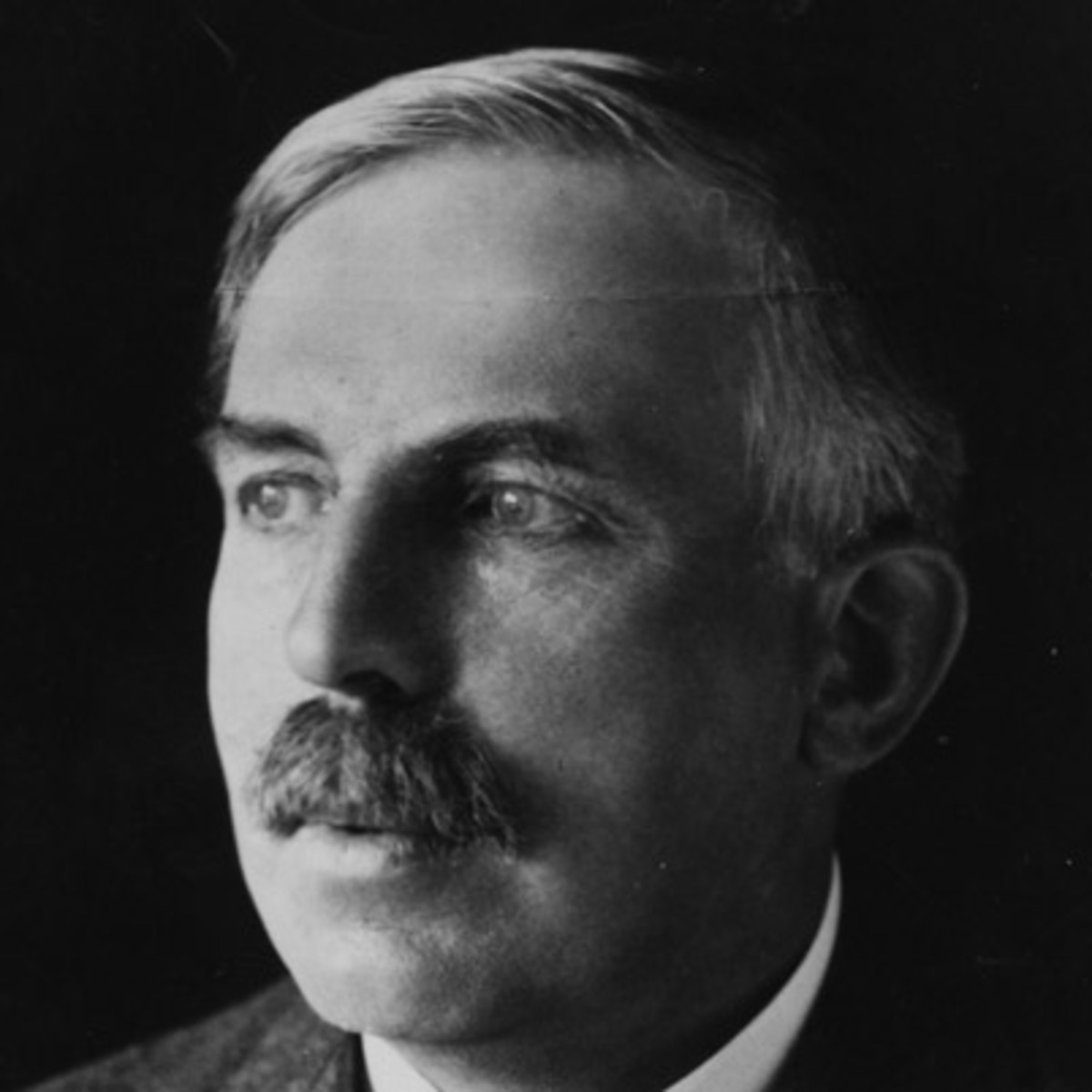}}
\caption{\small
Ernest Rutherford.}
\label{figu6}
\end{figure}

Ernest Rutherford (1871-1937) was born in Brightwater, near Nelson, New Zealand into a farmer's family in which he was one of 11 siblings. In 1895, he went to England to continue his education at the Cavendish Laboratory, University of Cambridge.  Rutherford (see Fig.~\ref{figu6})  was among the first of the ``aliens" (those without a Cambridge degree) allowed to do research at University of Cambridge.

In early work, Rutherford discovered the radioactive element radon, introduced the concept of radioactive half-life, and demonstrated the difference between  $\alpha$ ($^4$He)  and $\beta$ ($e^\pm$) decays of radioactive elements. This was the basis for the 1908 Nobel Prize in Chemistry  awarded  to Rutherford ``for his investigations into the disintegration of the elements, and the chemistry of radioactive substances."

In 1909 he suggested the Geiger-Marsden experiment on scattering of alpha particles passing through a thin gold foil. A breakthrough result was the observation of alpha particles with very high deflection angles while the expected distribution was homogeneous. Rutherford interpreted the data in 1911 by formulating the Rutherford model of the atom  according to which a very small positively charged nucleus, containing much of the atom's mass, was orbited by low-mass electrons.

In the 1917 experiment in which  nitrogen nuclei were bombarded with $\alpha$ particles Rutherford  discovered the emission of a subatomic particle which he called proton in 1920. 

\begin{figure}[h]
\epsfxsize=6cm
\centerline{\epsfbox{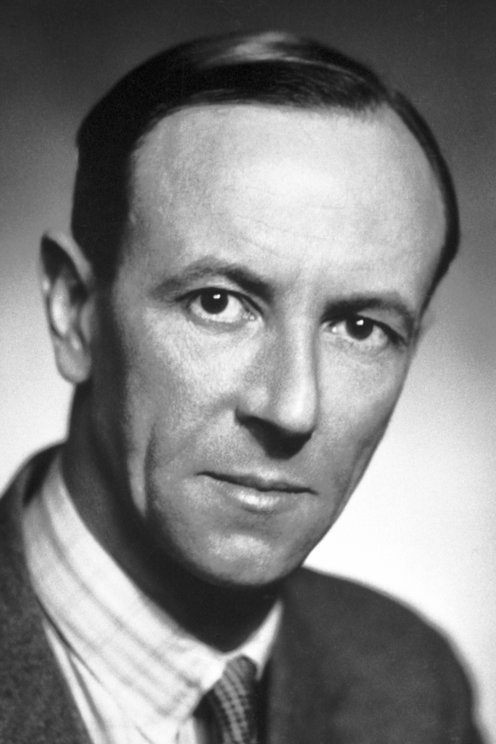}}
\caption{\small
James Chadwick.}
\label{figu7}
\end{figure}

James Chadwick (1891-1974) a British physicist who discovered neutrons worked at the same Cavendish laboratory under the directorship of Ernest Rutherford. He was born into the family of a cotton spinner and a domestic servant. After finishing primary school he was offered a scholarship to Manchester Grammar School, which his family had to turn down as they could not afford the small fees that still had to be paid. At the age of 16, he sat two examinations for university scholarships, and won both of them. At the University, Chadwick meant to study mathematics, but enrolled in physics by mistake.

In January 1932, Chadwick's attention was drawn to a  surprising result obtained by  Fr\'ed\'eric and Ir\`ene Joliot-Curie\,\footnote{Ir\`ene Joliot-Curie was the daughter of Pierre and Marie Curie.} 
who believed that they had managed 
to knock protons out of paraffin wax using polonium and beryllium as a source for what they thought was $\gamma$ radiation.  Upon reflection Chadwich came to the conclusion that their statement was wrong -- $\gamma$ quanta did not have enough energy to knock out protons. Neutrons (which were not yet discovered at that time) could do the job.

Immediately,  Chadwick focused on proving the existence of the neutron. He devised a simple apparatus that consisted of a cylinder containing a polonium source and beryllium target. The resulting radiation could then be directed at a material such as paraffin wax; the displaced particles, which were protons, would go into a small ionization chamber where they could be detected  with an oscilloscope.
In February 1932, after only about two weeks of experimentation with neutrons, James Chadwick sent a letter to ``Nature" reporting evidence of neutron's existence.  His discovery of the neutron completed the general picture of the atomic nucleus. The stage was set for the beginning of nuclear age.

For his discovery of the neutron, Chadwick was awarded the Nobel Prize in Physics in 1935. His discovery made possible artificial production of trans-uranium elements (i.e. heavier than uranium). This inspired Enrico Fermi to investigate the nuclear reactions induced by collisions of slow neutron with nuclei, work for which Fermi would receive the Nobel Prize in 1938.

In 1941, James Chadwick wrote the final draft of the MAUD Report, which gave a crucial  boost to the U.S. government launching the Manhattan Project during the Second World War -- atomic bomb project to which I will return later.

\section{History}

{My narrative in Sec. 3 is an adaptation of \cite{shif2}.}

\subsection*{\small Rudolf Peierls}

\begin{figure}[h]
\epsfxsize=8cm
\centerline{\epsfbox{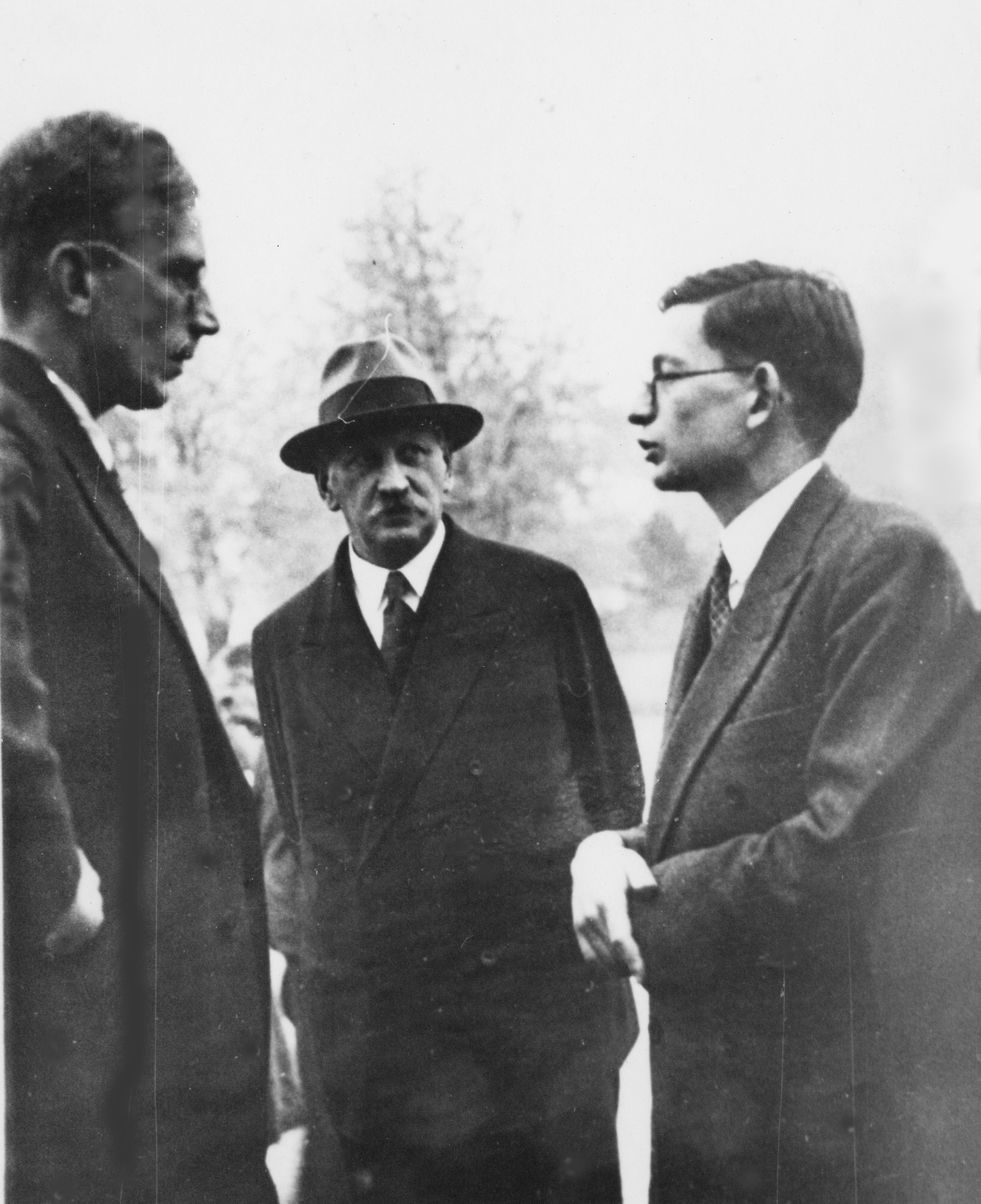}}
\caption{\small
George Gamow, Abram Ioffe, and Rudolf Peierls.}
\label{figuk0}
\end{figure}

On August 19-24, 1930, the All-Union Congress of Physicists was held in Odessa. This was the first physics conference organized in the USSR after the beginning of WWI. In attendance were over  800 delegates.
Foreign guests
were also invited, from the venerable -- Arnold Sommerfeld, Walther Bothe,
Wolfgang Pauli, and Franz Simon,\footnote{Franz Simon (Sir Francis Simon, 1893-1956), was a German and later British physical chemist and physicist who devised the gaseous diffusion method of separating the isotope $^{235}$U  and thus made a major contribution to the creation of the atomic bomb.
Friedrich ``Fritz" Houtermans (1903-1966)  was a  German nuclear physicists. In 1929, he made the first calculation of stellar thermonuclear reactions (with Robert d'Escourt Atkinson). 
After WWII he was one of the founders of nuclear geochronology.}  to the young -- Rudolf Peierls
and Fritz Houtermans. At that time Peierls was Pauli's assistant (see Fig. \ref{figuk0}).

Peierls who is the rightmost in this figure was 23 years old and did not know yet  that Providence had chosen him to place at the center of the nuclear saga. At this conference he met his future wife Genia Kannegiser. She had just graduated from Physics Department of Leningrad University. They spent a week together at the  conference and another week on board of {\em Gruziya}, a ship which took most of the participants of the event for a sea  excursion from Odessa to Crimea and then to Batumi in Soviet Georgia. Rudolf and Genia fell in love, and after Rudolf left for Zurich they started 
corresponding almost on a daily basis. In a few months Peierls learned Russian, and they switched from English to Russian in their letters.
This romantic year-long correspondence
 survived \cite{lee}. I advise you to look through it -- you will hardly find anything similar in modern life.

\begin{figure}[h]
\epsfxsize=10cm
\centerline{\epsfbox{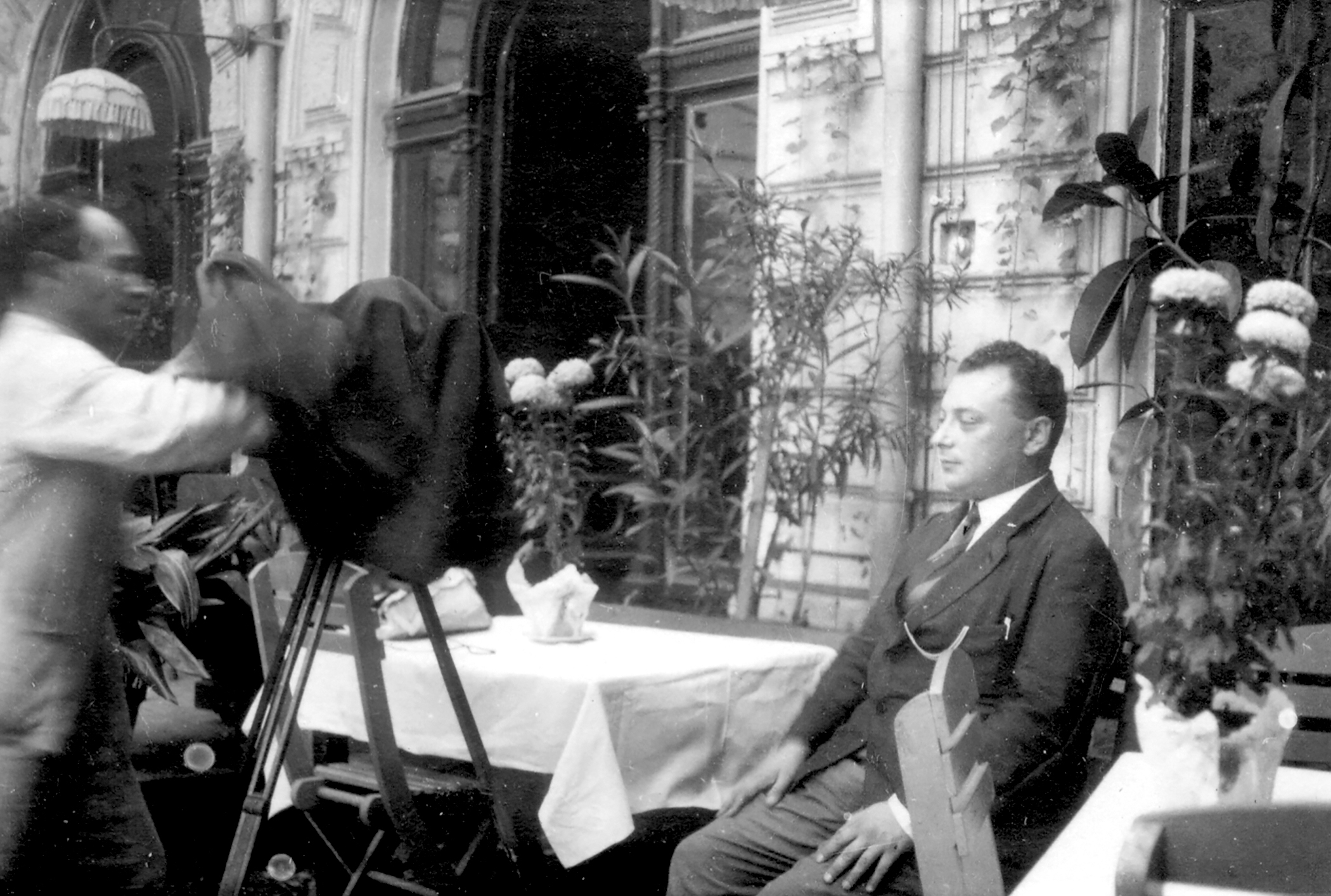}}
\caption{\small
Pauli in Odessa, August 1930 (courtesy of Pauli Archive at CERN).}
\label{figuk00}
\end{figure}

In 1931, Peierls came to Leningrad to deliver (in Russian!) lectures on quantum theory of condensed matter in which he was deeply immersed while in Zurich. And this is despite Pauli's well-known aversion to condensed matter physics! One had to have a strong character to do this. 

Rudolf and Genia married. In six months Genia managed to obtain permission to leave the USSR. This was a  miracle. Their marriage lasted for 55 years, until Genia's death.

Peirels' assistantship with Pauli came to an end in 1932. At that time getting a job for young physicists was extremely difficult -- much more difficult than now. Of course the number of theorists in the 1930s was incomparably smaller than now, but the number of open positions was smaller still. Post-doctoral positions did not exist, nor government scholarships. There were only two options: to become an assistant to a famous professor or to won a Rockefeller Foundation fellowship (for one year). In 1932, Peierls was one of three theoretical physicists in the world to receive it. The Rockefeller Foundation is the same foundation that gave scholarships to Landau, Gamow, Bethe, and dozens of other bright young physicists of that generation.  Then quantum physics was considered the most important science. The Foundation still exists, but, alas, physics has lost its place on the pedestal.

In the fall of 1932 the Peierlses arrived to Rome. Rudolf's idea was to work with Enrico Fermi to broaden his horizons. And indeed, Fermi was about to start planning experiments on bombarding uranium with neutron beams. Rudolf, increasingly drawn to nascent nuclear physics, decided to split the scholarship into two parts: six months in Rome, with Fermi, and six months in Cambridge, at the most prestigious university in England. Enrico Fermi was a recognized leader in the nuclear field, both an experimenter and a theorist. This still occurred at that time, but was extremely rare.

In December of 1932 Peierls received an offer of a position from Wilhelm Lenz (1888-1957), a professor at Hamburg University. Physics-wise he was not distinguished, but he was known for exceptional assistants. For instance, Ernst Ising\,\footnote{Ernst Ising (1900-1998 ) was a German physicist,  best remembered for the development of the celebrated Ising model. } was one of them, and then Wolfgang Pauli!

After a few euphoric days and a long discussion with Genia, Rudolf Peierls decided to decline the offer. This was a risky step, but also a wise one. The National Socialist German Workers' Party, commonly referred to  as the Nazi Party, had just won parliamentary elections in Germany. The advent of the Nazi government was imminent. Having experienced the Stalin and Mussolini regimes the Peierlses  anticipated that the Hitler regime would not be better.

In April 1933 the Peierlses  moved to Cambridge. This was a six-month visit -- not much time to start new projects.  Peierls established good working contacts with Paul Dirac. You won't believe, during war time Dirac worked with Peierls on uranium isotopes separation!

Time flew, the Rockefeller fellowship term was about to end with no job in sight.
Peierls was relieved when Lawrence Bragg\,\footnote{Sir William Bragg  (189 -1971) was a British physicist and X-ray crystallographer, discoverer (1912) of Bragg's law of $X$ ray diffraction. He received  the 1915 Nobel Prize in Physics.} arranged a scholarship for him through an Academic Assistance Council which 
helped German refugees in England. Bragg headed a crystallography group at Manchester University.

Continuing his condensed matter explorations Peierls paid more and more attention to nuclear physics. The transition was facilitated by a lucky coincidence: his old friend 
Hans Bethe was also in Manchester and even resided in the Peierlses' house.  Peierls and Bethe  would often refer to their  Manchester
time  as one of the most enjoyable and productive in their 
careers. In addition to starting considering nuclear reactions in stars (stellar nucleosynthesis), they developed, in particular, the theory of 
deuteron photo-disintegration. 

 In a conversation with James  Chadwick in Cambridge, Chadwick had challenged the two
friends to develop a theory of this phenomenon. On the train back
from Cambridge to Manchester, a journey of about four hours, 
after intense conversations with the help of numerous backs
of envelopes, they succeeded in solving the problem. You can find their solution in any good quantum mechanics textbook.
Peierls also  worked
on Dirac's hole theory and on the newly introduced Pauli's neutrino.

In 1936, Mark Oliphant\,\footnote{Sir Marcus Oliphant (1901-2000) was an Australian-British physicist  who played an important role in the development of radars and the uranium isotopes separation program.}  was appointed head of the Department of Physics at the University of Birmingham. He had to finish his business in Cambridge and therefore agreed that he would move to Birmingham in October 1937. In the spring, Mark approached Rudy and asked: ``What would you say if I tried to organize for you a chair of theoretical physics in Birmingham?"

In almost all English universities, theoretical physics was not considered a separate science. Only enthusiasts at the departments of applied mathematics were engaged in theoretical physics, but in fact it was mathematical physics, only indirectly related to experiments on quantum phenomena, which, in fact, determined the face of the then ``new" physics. Theoretical physics associated with experiment was Rudolf's dream. Of course, he agreed.

The Peierlses arrived in Birmingham in October of 1937. By 1938 or so his transformation to a nuclear theorist was complete. Its culmination was the Frisch-Peierls Memorandum which we will discuss on pages \pageref{fpm}, \pageref{memom}.

\begin{figure}[h]
\epsfxsize=11cm
\centerline{\epsfbox{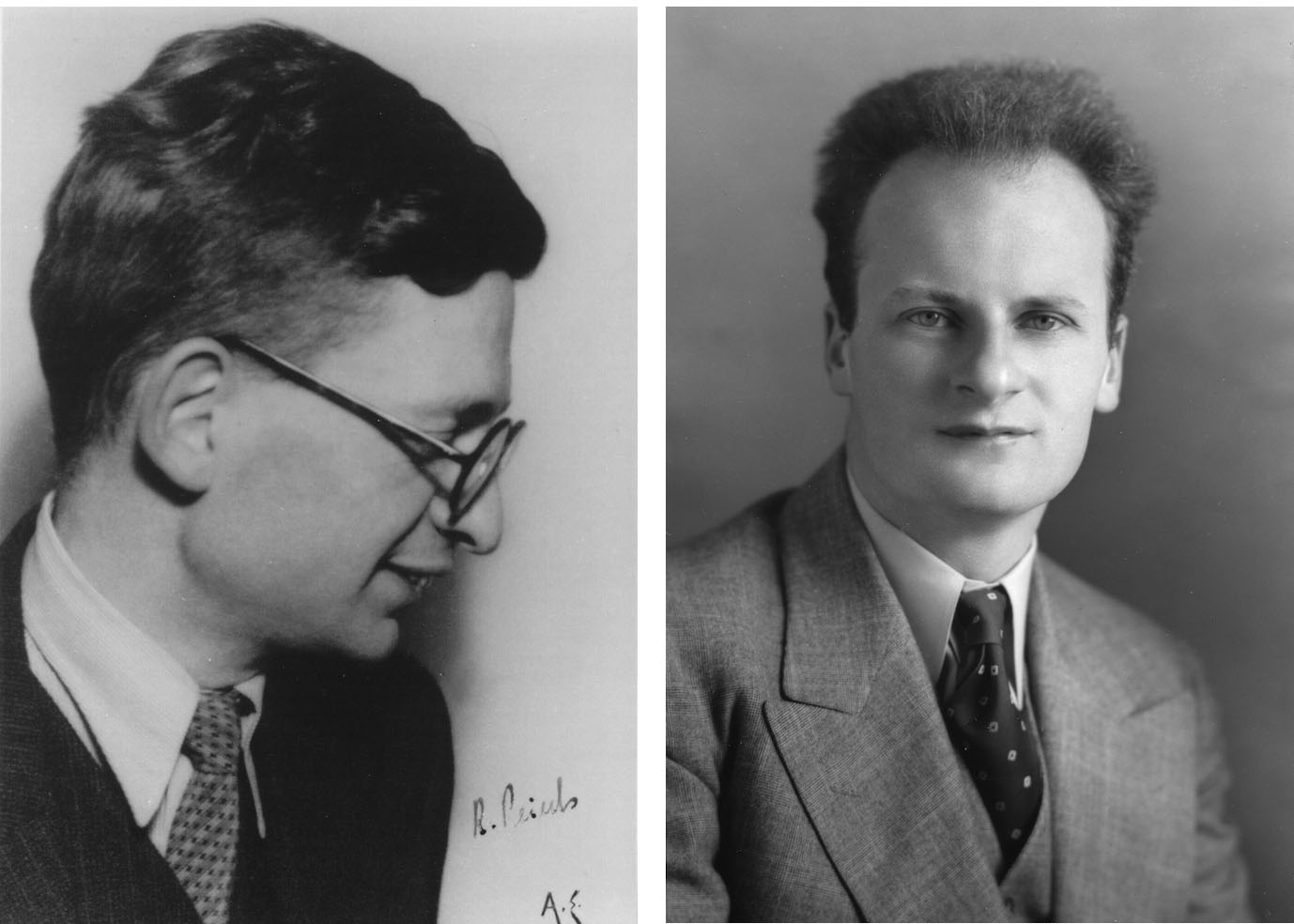}}
\caption{\small
Rudolf Peierls and Hans Bethe in the late 1930s.  Courtesy of AIP Emilio Segr\`e Visual Archives, Simon Collection.}
\label{figuk00}
\end{figure}

\subsection*{\small Lise Meitner and Otto Frisch}

In 1934, Lise Meitner (1878-1968), then working at the Kaiser Wilhelm Institute in Berlin, persuaded her longtime colleague, the renowned chemist Otto Hahn (1879-1968), to start a group for neutron irradiation of uranium and study of the resulting products. Meitner, Hahn and Strassmann\,\footnote{Fritz Strassmann (1902-1980) was Hahn's young assistant.} quickly repeated Fermi's experiments and went further. Until 1938, it was generally accepted that neutrons, penetrating into a heavy nucleus, were held there in ``captivity", giving life to still heavier nuclei, hitherto unknown (trans-uranium).   This is exactly what Fermi thought after his experiments in Rome in 1934. By 1938, Meitner, Hahn, and Strassmann had published a dozen papers supporting the transuranium hypothesis.

\begin{figure}[h]
\epsfxsize=11cm
\centerline{\epsfbox{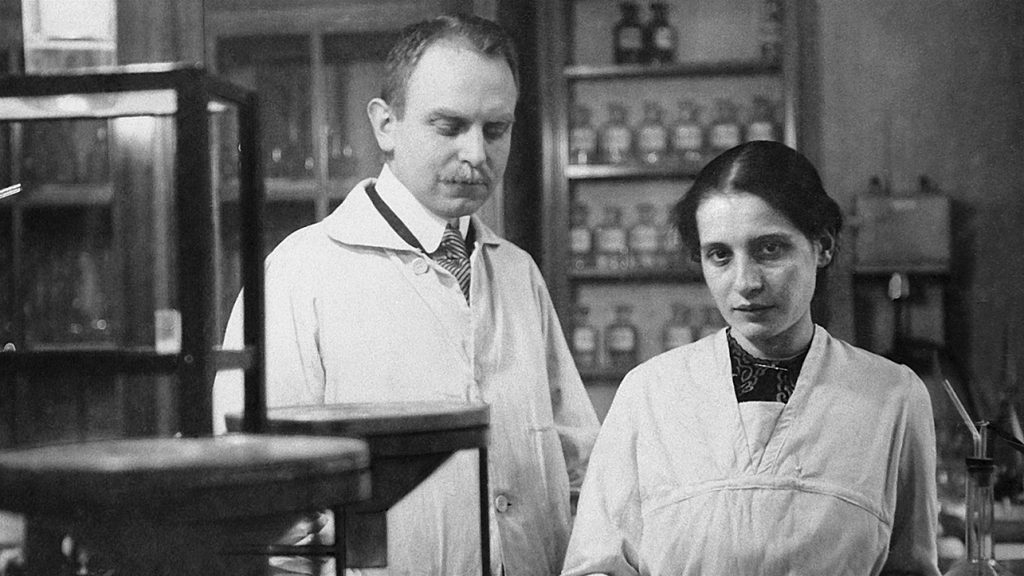}}
\caption{\small
Right to left: Lise Meitner and Otto Hahn.}
\label{figu8}
\end{figure}

And yet something was bothering Lise. In the products of irradiation of the uranium target, chemical elements lighter than uranium, the origin of which she could not understand, showed up. In early July 1938, Otto Hahn informed Lise that he had again carefully analyzed the products of the previous experiment and found three substances in them that chemically behave like radium (atomic number 88) -- four atomic units below uranium. ``I decided to do a new analysis after I received a letter a week ago from Ir\`ene Curie and Pavel Savich,\footnote{Pavel Savi\'c (1909-1994) was a Serbian physicist and chemist. In the years 1937 and 1938, he worked with Ir\'ene Joliot-Curie and Fr\'ed\'eric Joliot-Curie on interactions of neutrons in chemical physics of heavy elements.}  who also mention radium," Hahn wrote.
Lise didn't have time to answer him right away. Other thoughts plagued her that day.
Back in 1937, Germany passed a law  according to which the Kaiser Wilhelm Institute became state property, with all the ensuing consequences. The few remaining Jews at the Institute were expelled, Lise last. She was safe (alas, not for long) as long as she was an Austrian citizen. In March 1938,  Hitlerannexed Austria, and all Austrians automatically became German citizens. Lise Meitner's position turned disastrous in just a day. Her dossier came to the consideration of Heinrich Himmler, who wrote: ``To fire, but not to let go of Germany ... It is highly undesirable for famous Jews to leave Germany; they should not be able to talk abroad about their attitude towards Germany."  Lise found out about this in June 1938. She was threatened with deportation to  concentration camp. Legal emigration was out of question.

When Niels Bohr learned about the mortal danger Lise was in, he immediately called his Dutch colleague Dirk Koster: ``What can we do for Lise?" Koster immediately arranged her entry visa to Holland. Solidarity of honest people ... 

Without permission to leave, she had to cross the German border illegally. In Berlin, only Otto Hahn and the Springer publishing house scientific advisor, Paul Rosebaud,\footnote{Paul Rosebaud also was the British intelligence agent in Nazi Germany.} were initiated into the planned escape.
Koster arrived in Berlin on Monday night. On Tuesday, July 12, Meitner came to the Institute early. Hahn told her about the details of Koster's plan.  ``Fraulein Meitner," he said, ``take this diamond ring. It once belonged to my mother. If you need to bribe the border guards, use it."
The next day, Rosebaud drove Lise to a Berlin train station. In the last minutes, already on the platform, fear paralyzed Lise. She clung to Rosebaud and begged him to take her home. Rosebaud refused. Koster was waiting for Lise on the train; they greeted each other as if they had met by chance. Nothing remarkable happened on the trip. As they approached the Dutch border, Lise became nervous, but the border crossing proceeded without incident.

After informing Hahn that she was safe, Lise wrote a letter to him: ``It seems to me that in the current situation, when we do not understand the origin of light nuclei, we should not publish our latest result until absolutely everything is clear." Hahn asked Strassmann to prepare a new series of experiments.
Soon Lise moved to Sweden.

She was saved from depression only by letters from Hahn, who, together with Strassmann, continued the experiments in Berlin. Otto consulted with Lise, discussed new results with her and asked her opinion on key issues. Collaboration with Hahn continued at the same pace as before her escape from Berlin. In November 1938, they met secretly in Copenhagen, where they discussed and charted subsequent measurements.

Usually Lise spent Christmas vacations in Berlin with her nephew Otto Frisch. Frisch's mother (Lise's sister) was a famous pianist. Like Lise, Otto was from Vienna. A favorite city to which they could not return ... Niels Bohr sheltered Otto at his institute in Copenhagen. That year, for the first time in many years, Lise and her nephew could not meet in Berlin and decided to spend their holidays in the small Swedish town of Kungalv near Gothenburg.

\begin{figure}
\epsfxsize=13cm
\centerline{\epsfbox{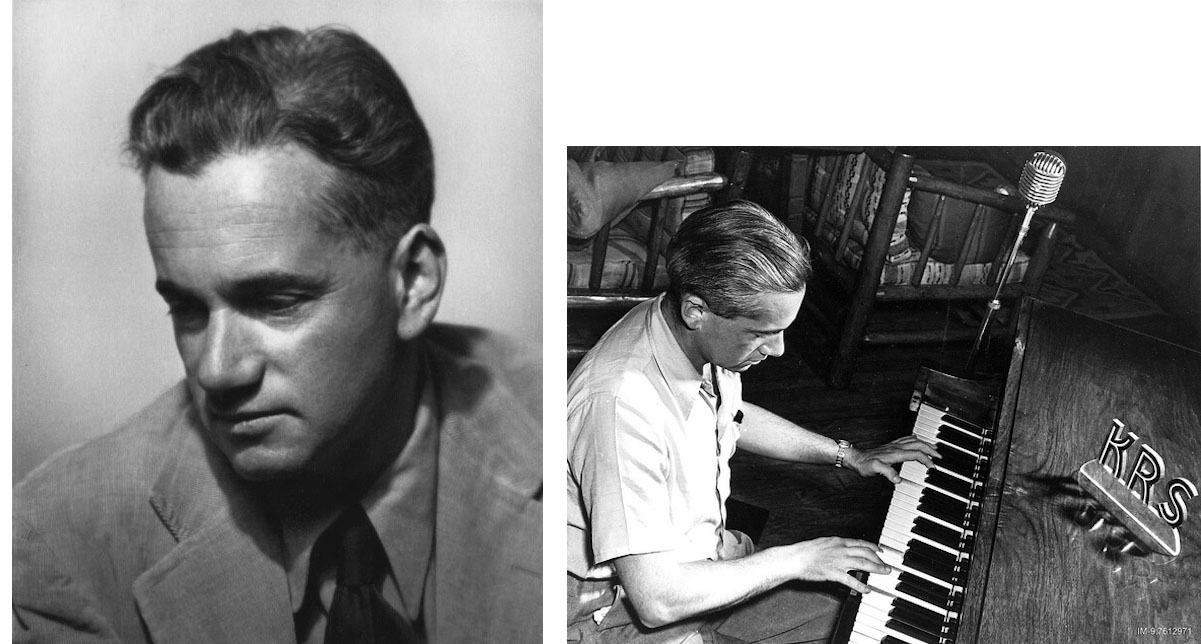}}
\caption{\small
Otto Frisch.}
\label{figu9}
\end{figure}

Frisch arrived in Kungalv a couple of days before Christmas late in the evening. In the morning, leaving the room and heading for breakfast, he saw Lise, immersed in reading a letter from Hahn. Lise didn't even hug or listen to her nephew. The first thing she did was hand him a letter.
Frisch was so impressed by its contents that he didn't believe his eyes and read it again. Hahn wrote that the three substances that they previously considered radium from their chemical signature are most likely isotopes of barium. But barium is not just slightly lighter than uranium -- its atomic number is 56, it is almost twice as light!

-- An error?

-- No, Otto, Hahn and Strassmann are too good chemists to make such a mistake.

-- But where does barium come from after irradiating uranium with neutrons? 

The conversation soon continued in the snowy forest -- Otto on skis, Lise keeping up with him. Word by word, they remembered Bohr's lectures, in which he drew a nucleus in the form of a liquid droplet. But what if a neutron, getting into the nucleus, does not get stuck in it, but divides this droplet into two parts? Is it possible? 

Otto and Lise sat down on the trunk of a fallen tree and figured out the energy balance on a piece of paper: it all came together. The electric repulsion of the protons almost compensated for the surface energy of the nucleus, so even a relatively weak ``slap" received from the neutron could split the uranium nucleus into two or even three parts. Moreover, Lise remembered the formula for the masses of nuclei -- right there in the forest they calculated the energy released in this process. Lise and Otto looked at each other -- the energy was enormous.

A couple of days later, Frisch left for Copenhagen to report -- how impatient he was! -- to Bohr about the discovery. Bohr was about to sail to America. He was very busy, but  listened to Frisch. Not even a minute had passed when Bohr hit himself on the forehead with his palm, exclaiming: ``We are   idiots! Awesome! This is how it should be. Is your note with Lise ready?" Frisch replied that they were starting to write a paper, and Bohr promised not to tell anyone until it was ready. With these words, he got into a taxi and went to the port so as not to be late for the ship.

Further events developed rapidly. Lise in Stockholm checked all the calculations again. On January 6, 1939, Hahn and Strassmann published an article in the German journal {\em Naturwissenschaften} in which they reported the discovery of barium after irradiation of uranium. Lise was not listed in the authors!

She was crossed out  although this was the last step in the long-term collaboration of the three authors. Hahn later said that even if the  Meitner had been included in the list of authors, it would still have been removed by the editorial office of the journal.

In Copenhagen, Georg Placzek convinced Frisch that their -- Lise and Otto's -- paper would benefit greatly from observing fast-moving uranium fragments. The experiments were then completed in  several days  -- what a drastic diffeence with experiments of today. In two days, Frisch assembled the installation, took measurements and actually found the necessary fragments. It took several more days to agree on the article by phone. On January 14, the article by Frisch and Meitner was mailed to London, to the journal {\em Nature}, where it was received on January 16, and  published  on February 11. Before completing their paper, Frisch asked an American biologist who happened to visit Bohr's nstitute: ``What word do you use for the process of amoeba division?" -- ``Fission." -- ``Great, that's exactly what I need. Now physicists will have their own fission."

\begin{figure}[t]
\epsfxsize=8cm
\centerline{\epsfbox{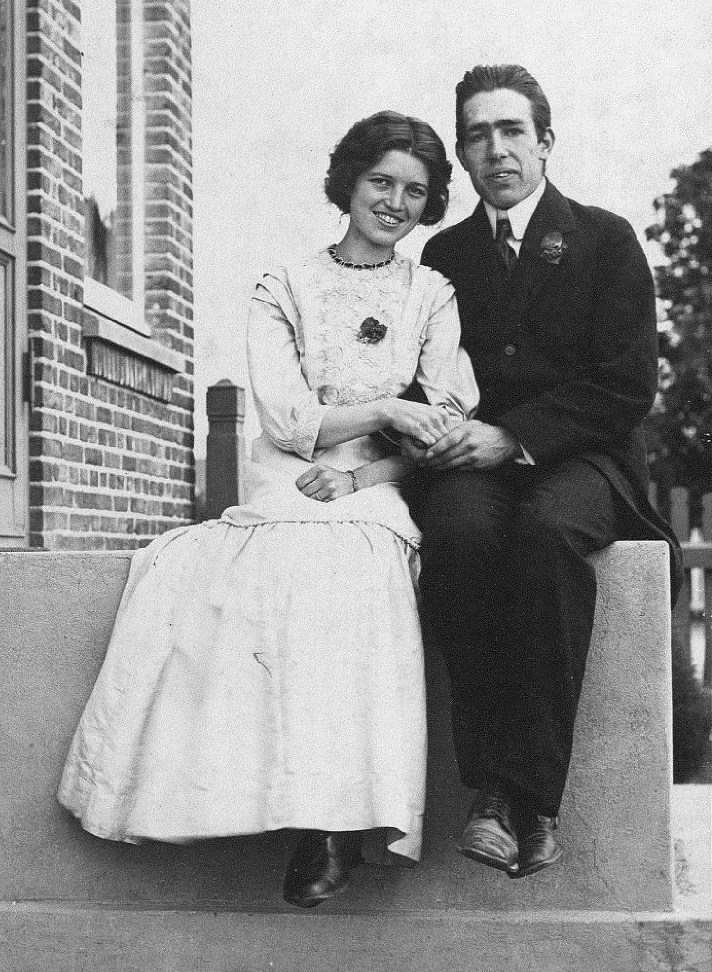}}
\caption{\small
Neils Bohr with his wife Margrethe, 1910.}
\label{figu10}
\end{figure}

So the first sight of a bomb loomed. Bohr reassured everyone: ``The secondary neutrons from the fragments are too fast to cause fission in natural uranium, which is 99\% uranium-238. They could cause fission of uranium-235, but there is too little of it. Bohr was perfectly right, but he did not take into account one circumstance: the unprecedented efforts and ingenuity of physicists and engineers, gathered in Los Alamos, Americans and Europeans, who were driven out by Hitler from the burning Europe. They were driven by a common feeling -- to prevent the world domination of Nazi Germany. They were driven by the fear that the army of the Third Reich would be the first to acquire nuclear weapons.\footnote{The uranium project was launched in Germany on September 1, 1939, simultaneously with the beginning of WWII. Werner Heisemberg assumed the role of the Head of this project.}

Concluding this section I should mention that Otto Hahn was awarded  the 1944 Nobel Prize for Chemistry (with the citation ``for his discovery of the fission of heavy nuclei") while Lise Meitner never received it despite protests from Niels Bohr and other distinguished physicists of this time. 
That was a sad time when female physicists were not valued. Moreover, some historians say that the Nobel committee in 1944 was pro-German and anti-Jewish. 
There is a nice 2006 PBS documentary titled {\em The Path to Nuclear Fission, The Story of Lise Meitner and Otto Hahn} which is
available on Youtube \cite{yt1}. A documentary on Niels Bohr presenting many rare and highly interesting documents was produced in 1985, unfortunately in Danish
\cite{yt2}.

\subsection*{\small Frisch-Peierls Memorandum}

\label{fpm}

In January 1940 Rudolf Peierls at Birmingham University  received a letter from Perrin,\footnote{Jean Baptiste Perrin (1870-1942) was a French physicist based in Paris who, in his studies of the Brownian motion of minute particles suspended in liquids, verified Albert Einstein's explanation of this phenomenon and confirmed the atomic nature of matter. For this achievement he was honored with the 1926 Nobel Prize for Physics.}  who wrote about calculations of  uranium critical mass needed for the explosion. In 1940 no experts in this issue existed. A handful of people were involved in nuclear fission, and everyone made mistakes. Rudolf Peierls made up his mind  to fix Perrin's mistake. He wrote a short note, but did not send it to the journal. He was tormented by the thought whether his modest result might help the Germans in the development of their nuclear weapons. Werner Heisenberg, the founding father of quantum mechanics, has suspiciously disappeared since the beginning of the war.
Peierls had first-hand information about the publications of Hahn, Meitner and Frisch. Frisch arrived in Birmingham in July 1939
to assume a one-year position of a university lecturer.  There was no better choice. The war began. Frisch could not return to Bohr in Copenhagen.

So, in February 1940, Peierls walked into Frisch's office with his unsent manuscript in hand. They scanned the draft together. ``I do not see anything that could cause the postponement of publication, Professor Peierls, -- said Otto, -- after all Bohr showed that the atomic bomb is not realizable. It will take tons of uranium, but even so, its size will be so large that the chain reaction will die out. There is nothing new on this issue  in Perrin's paper, nor in yours. Submit it to the journal."

\begin{figure}[t]
\epsfxsize=10cm
\centerline{\epsfbox{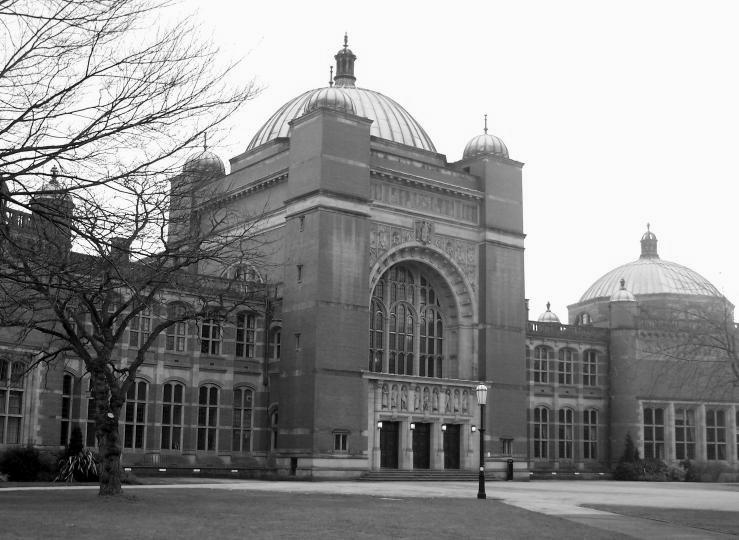}}
\caption{\small
Birmingham University.}
\label{figu11}
\end{figure}

A few days later, Frisch nocked at Peierls's office door and asked: ``What if we consider not uranium in general, but only its lighter isotope, uranium-235?"

The work took only a few days. It often happens when a new horizon suddenly opens up. All the elements of the mosaic, stored in the depths of consciousness, suddenly combine into a single integral picture. Here is the last page of calculations. Frisch wrote in large letters: ``The critical mass for uranium-235 is about 600 grams" -- and outlined it with a bold frame. Frisch and Peierls  looked at each other, stunned, and froze. A couple of pounds instead of a ton! (In 1942, the US physicists more accurately measured the cross section  of interaction of neutrons with uranium-235 nuclei and updated  the prediction for the critical mass for uranium-235. It grew to several kilograms). Then  Frisch and Peierls quickly estimated the total 
energy release and looked at each other again in awe of the destructive force that opened before their eyes.

``Even if it costs as much to build an isotope separation plant as a battleship, in order to end the war once and for all, it must be done. And God forbid, in Germany it will be done before us. The British government should be informed immediately," summed up Peierls.
They decided to write a memorandum, and try to keep its content secret, so that not a single word would reach the physicists who remained under Hitler. By mid-March 1940, a memorandum was ready. It is now known as the Frisch-Peierls Memorandum. It can be found in any textbook on the history of physics, it begins as follows:

\begin{center}
\subsection*{\small Memorandum: On the Construction of a
``Super-bomb" based on a Nuclear Chain
Reaction in Uranium}
\end{center}
\label{memom}

\begin{quote}
The possible construction of ``super-bombs" based on a
nuclear chain reaction in uranium has been discussed
a great deal and arguments have been brought forward
which seemed to exclude this possibility. We wish here to
point out and discuss a possibility which seems to have
been overlooked in these earlier discussions. [...]

The attached detailed report concerns the possibility of
constructing a ``super-bomb" which utilizes the energy
stored in atomic nuclei as a source of energy. The energy
liberated in the explosion of such a super-bomb is about
the same as that produced by the explosion of 1,000 tons
of dynamite. This energy is liberated in a small volume,
in which it will, for an instant, produce a temperature
comparable to that in the interior of the sun. The blast
from such an explosion would destroy life in a wide area.
The size of this area is difficult to estimate, but it will
probably cover the centre of a big city. [...]

In addition, some part of the energy set free by the bomb
goes to produce radioactive substances, and these will
emit very powerful and dangerous radiations. The effects
of these radiations is greatest immediately after the
explosion, but it decays only gradually and even for days
after the explosion any person entering the affected area
will be killed.

Some of this radioactivity will be carried along with the
wind and will spread the contamination; several miles
downwind this may kill people.[...]
\end{quote}

In May 1940, Winston Churchill became Prime Minister. In direct response to the Frisch-Peierls memorandum, one of his first acts was the creation of  a committee to consider the possibility of making A-bomb.
The initial committee called MAUD committee confirmed  conclusion by Frisch and Peierls.
After fifteen months of work, in summer of 1941 it issued a report ``Use of Uranium for a Bomb"  known as the MAUD Report. 
After examining the latter, the British Government judged the bomb to be high priority.  Churchill asserted, ``Although personally I am quite content with the existing explosives, I feel we must not stand in the path of improvement, and I therefore think that action should be taken..."  
At that time  a nuclear weapons project officially named Tube Alloys was set in motion. ``Tube Alloys" was charged with working out technical details
necessary for the implementation of the project.

When President Roosevelt wrote to Churchill in August 1941 suggesting collaboration, Churchill responded unenthusiastically, preferring to keep the more advanced Tube Alloys project separate. In 1941 Americans were far behind. A year later the situation dramatically changed. The MAUD Report which was made available to the United States energized the American effort. Now Churchill wanted to combine the British and American efforts.  
In 1943 the fusion of Tube Alloys with the Manhattan project became inevitable. This was the beginning of Los Alamos and the beginning of the nuclear age.

\begin{figure}[h]
\epsfxsize=9.5cm
\centerline{\epsfbox{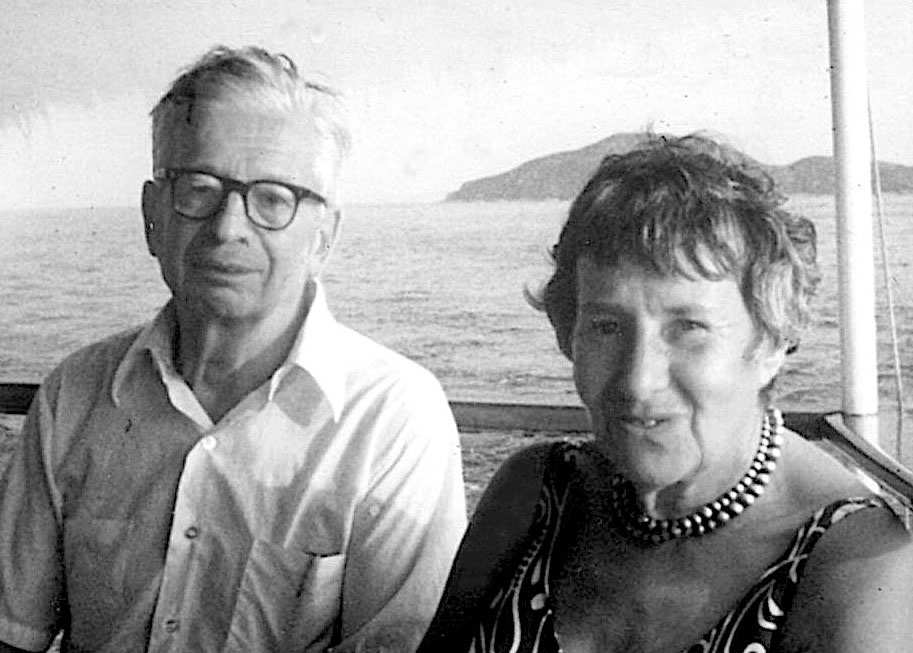}}
\caption{\small
Rudolf and Genia Peierls, 1979, courtesy of Natalia Alexander.}
\label{figu11a}
\end{figure}

\subsection*{\small Epilog}

Before finishing my narrative I have to make two remarks.

First, as you certainly know, the German nuclear project led by Heisenberg turned out to be a complete failure. Why?
The answer to this question was given by Peierls himself in \cite{ph}. I will quote a couple of paragraphs from this article. 

\begin{quote}

Once [Heisenberg] had decided to remain in Nazi Germany, he could hardly avoid making concessions to the regime. Not everybody could be like Max von Laue, who kept out of sight and made as few concessions as possible (it is said that he never left the house without carrying two packages or a briefcase and a package, so that he would have no hand free to give the Hitler salute). This was not a possible way of life for Heisenberg, who wanted to continue being active as a leader in science [...].

[Heisenberg and the other German] scientists overestimated the difficulties, because they never made a careful estimate of the critical size needed for a fast chain reaction, and therefore tended to overestimate the size of a weapon. It is also true that there were many errors of judgment in their work. For example, they excluded graphite as a moderator to slow down the neutrons because of a mistaken finding that it absorbed too many neutrons. This failure is often blamed on Walther Bothe,\footnote{See footnote \ref{bothe}.} who is alleged to have made ``wrong'' measurements. In fact his measurements were correct; but the ``pure'' graphite he was using contained impurities in amounts that were too small to be detected chemically, but were fatal for the absorption of neutrons. In the same situation, Fermi and his collaborators in the US guessed that further purification would improve the results. 
\end{quote}

It seems to me that the general (non-technical) reason for the failure is a huge difference in motivations. Unlike the physicists involved in the Manhattan project those few who stayed in Germany 
and worked under Heisenberg's leadership were not enthusiastic enough (and, perhaps, less talented). They lacked the feeling  that  their success was crucial for the survival of human civilization -- the feeling which ignited the hearts and minds of physicists on the other side of the Atlantic Ocean.

And my last remark is about 2014-2015 Amazon production titled ``Manhatten." In fact, it is about Los Alamos, over 20 episodes 
the only merit of which is that they are colorful. Everything else is twisted beyond any reasonable limits: Robert Oppenheimer is presented as  a boorish egocentric, 
Niels Bohr as an idiot, the British scientists as second-rate clowns, instead of camaraderie they (the series makers) show animosity, lots of sex, the moral cliches of today are projected onto Los Alamos of 1943, etc. etc.

\newpage

\vspace{0.5cm}
\subsection*{\small Appendix}
\vspace{0.3cm}

\begin{center}
{  \large \bf  Early Years of Quantum Mechanics}\footnote{The Russian original was published in \cite{rp}. I am grateful to A. I. Chernoutsan, Deputy Editor-in-Chief of {\sl KVANT},  
for kind permission to translate this article in English and publish the English translation. Unfortunately, the original Peierls's photographs donated by him to {\sl KVANT} and published in \cite{rp} cannot be found. I replaced them by similar photographs. If not stated to the contrary, they are taken from Wikimedia Commons. Figure 1 first appeared in my publication \cite{shif1}.  I am grateful to A. Varlamov for pointing out to me Peierls's talk published in \cite{rp}.

  }
\\[2mm]
RUDOLF PEIERLS

\end{center}

 \color{black}

Beginning this talk from a narrative about myself might seem immodest. Usually people do not do that.
But since I will speak about my personal impressions I have to introduce myself. I entered the university in 1925. Today I would want to be
able to say that I had chosen physics because it was an interesting subject and was rapidly developing.
However, this would not be fair. I actually wanted to be anengineer. That was the time when
aviation, new cars were being developed and it was natural that the boy wants to become an engineer. But 
they said that I was not fit for this, that I would not be a good engineer. Why? -- I do not know. Therefore
I chose, as it seemed to me, the subject closest to my dream -- physics.

I went to University of Berlin -- my home city. My parents thought I was too young to go far away. There I attended
lectures by Max Planck. These were the most nasty lectures which I had ever had.
His lectures followed his book verbatim. If you had a copy of this book with, you could follow  line by line. Planck was already very famous then but  we still had no clue as to where his fame came from. 
The first time I heard about Planck's constant, Bohr's atom and stuff like that was at Walter Bothe's lectures.\footnote{Walther Wilhelm Bothe (1891-1957) was a German nuclear physicist, who shared the Nobel Prize in Physics in 1954 ``for the coincidence method and his discoveries made therewith," with Max Born. \label{bothe}} (Later Bothe became nuclear physicist.)   Then it became clear to me that something new and very interesting was going on in physics.

In a year I decided that I was adult enough to leave Berlin. I moved to Munich, where Arnold Sommerfeld -- the best teacher of theoretical physics -- worked. It was  wonderful time for theoretical physics. Quantum mechanics was in the making. Today, it is hard to imagine how fast it occurred -- in fact, in two years.

Just then I enrolled at  University of Munich. In a year I could read papers on quantum mechanics. But I was late to catch its formative period. If I could repeat my life I would like to be born a year or two earlier. Felix Bloch\,\footnote{Felix Bloch (1905-1983) was a Swiss-American physicist and Nobel physics laureate who worked mainly in the U.S. He and Edward Mills Purcell were awarded the 1952 Nobel Prize for Physics for ``their development of new ways and methods for nuclear magnetic precision measurements."} later explained me that not every person is fit to create new theories and that we serendipiously appeared just in time to apply them. To my mind, he was right. It was the right time to pick up  a problem which ``old" physics  could not solve -- there were plenty of them -- and apply to them new methods.

\begin{figure}[t]
\epsfxsize=11cm
\centerline{\epsfbox{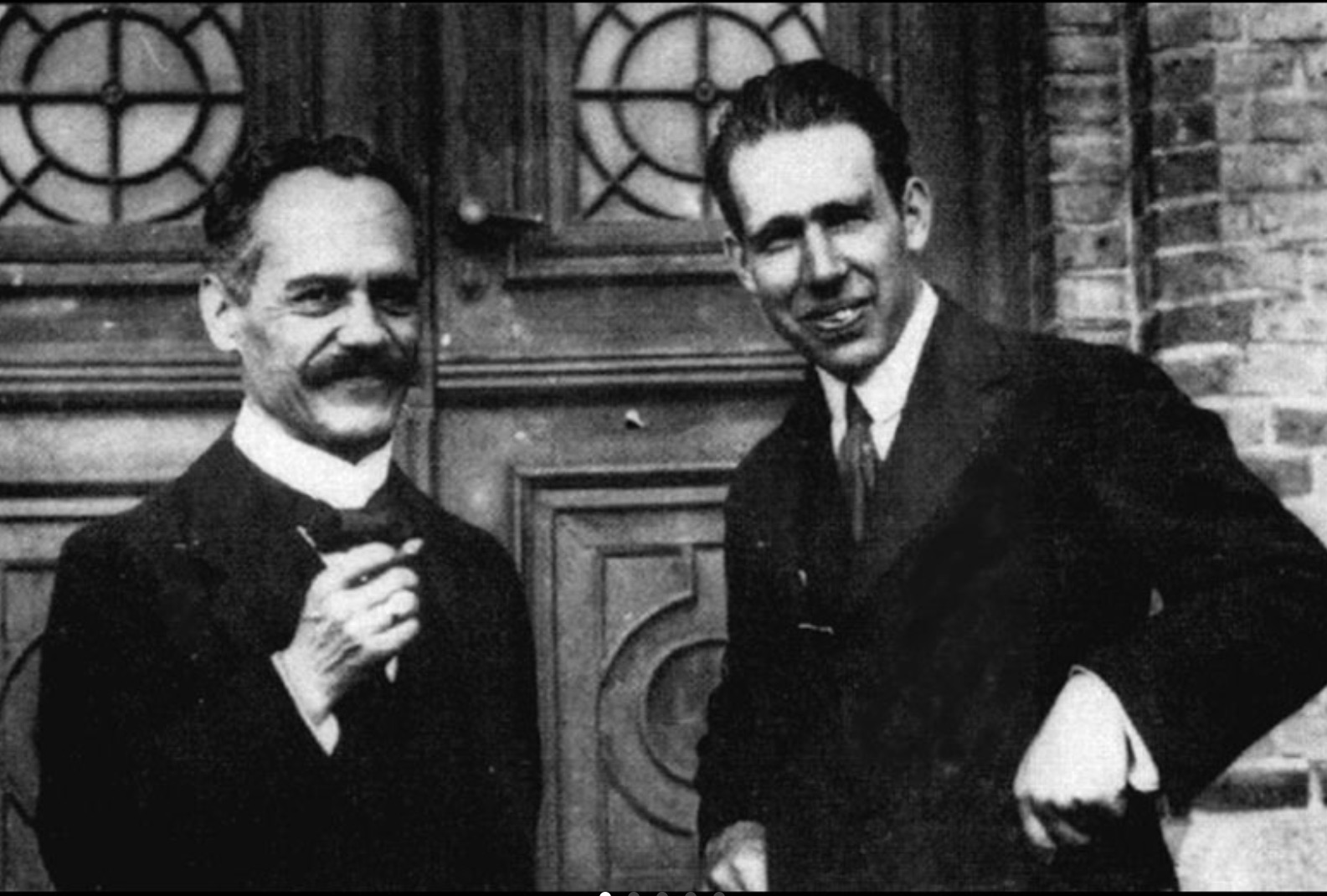}}
\caption{\small
Arnold Sommerfeld and Niels Bohr.}
\label{figuk1}
\end{figure}

Thus I came to Sommerfeld. He was short but  had a huge mustache. Sometimes we called him ``upper halh and a little bit more."
Sommerfeld
looked pretty important and had the title of {\em Geheimrat} -- Privy councilor. One can compare this title with the modern Russian Academician. He liked when people addressed him this way. An American student at first did not know about this and would address Sommerfeld ``Herr Professor." In a week or two his fellow student explained him the difference. Next time he met Sommerfeld he addressed him ``Herr Geheimrat." Sommerfeld noted the change and said: ``Your German has significantly improved lately."

But in our institute Sommerfeld was anything but Geheimrat. We never addressed him that way. He was a wonderful teacher both for undergraduate and graduate students, his lectures were of remarkable clarity. They are published and are not dated even today; it is instructive to read them. Sommerfeld  always insisted that theoretical physics as  science should be based on experimental data. He never let us forget what particular facts of nature laid in the foundation of this or that theoretical law.

\begin{figure}[t]
\epsfxsize=7.5cm
\centerline{\epsfbox{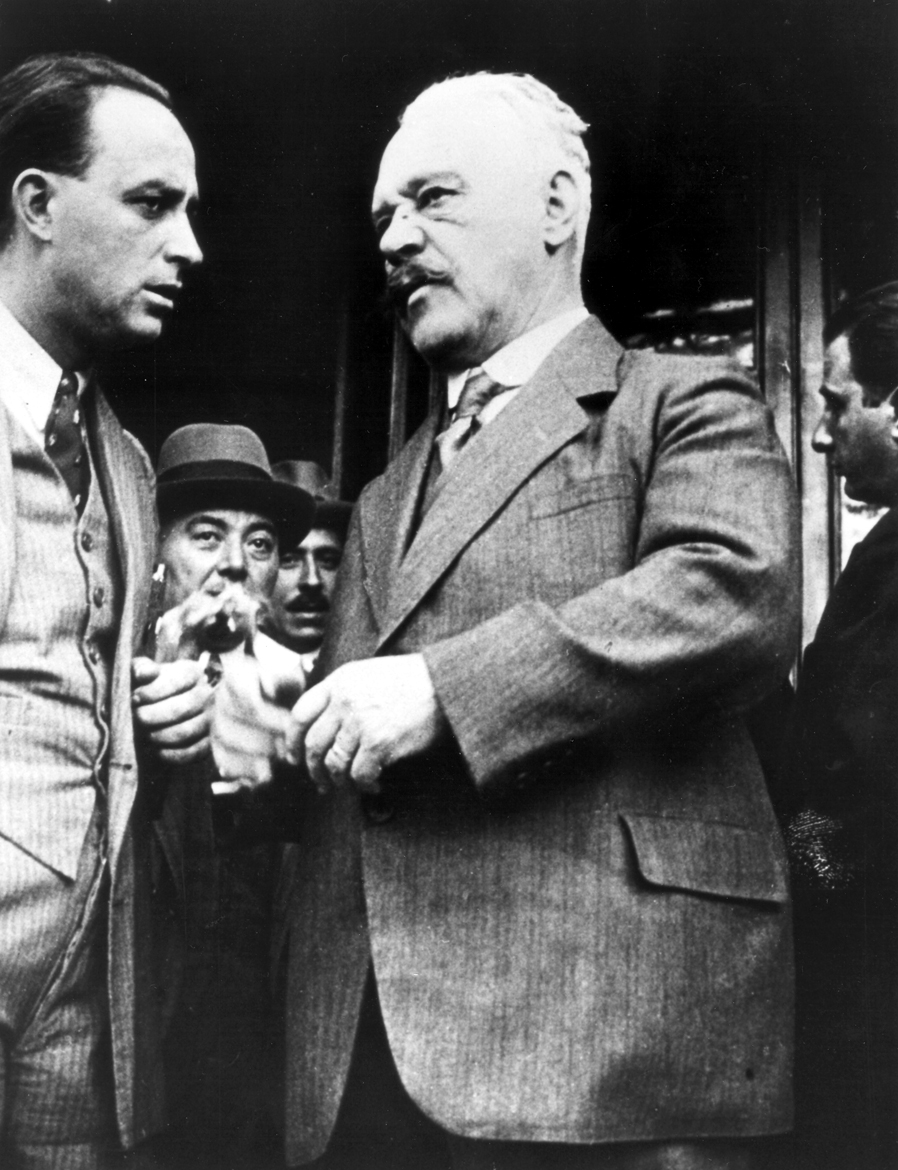}}
\caption{\small
Enrico Fermi and Arnold Sommerfeld.}
\label{figuk2}
\end{figure}

Sommerfeld's knowledge of mathematics was excellent. He published many purely mathematical papers -- very useful -- but never was too meticulous.
I remember his lectures on electron theory of metals. In a calculation he carried out in a blackboard he lost a factor of 2. We noted the omission which did not seem important to us. Finally, he approached the Wiedemann-Franz law in which the numerical coefficient is common knowledge. Sommerfeld stared at the blackboard realizing that he was going to obtain a wrong coefficient in this law. Noting the error, with no interruption, 
Sommerfeld spelled out: ``We should take into account both the left-moving and right-moving electrons, therefore the missing 2 should appear right here," and moved on.

Sommerfeld had a small vacation cabin in the mountains. Sometimes he invited there graduate students and other professors. My first seminar talk occurred in this cabin. He invited me.  At this time the works by Dirac and Jordan on theory of transformations just appeared. "We had no time to understand whether these works are important, -- said Sommerfeld, -- but perhaps you will be able to explain them to us."

This was a hard task for the student who had spent only two years at the university. However, I gladly agreed. I do not know whether other participants of the seminar learned something from my presentation but I learned a lot. 

At that time Hans Bethe was a graduate student at Munich. He was only one year older than me. In this age it is a big difference. He seemed to me a wise man from whom I could learn a lot. We made friends. He is still older than me by a year. Now it is not so important.I still can learn many things from him. I spent in Munich 18 months, and would gladly stay for longer. However, Sommerfeld was invited to the US for a year or so. Before leaving he advised me to go to Leipzig to work with Heisenberg.

Heisenberg was the opposite of Sommerfeld. 
I did not notice not even a trace of {\em Geheimrat}  there. 
In appearance, in any case, he was a very modest man. Typically, we had a seminar each week. Before the seminar a small tea party was always arranged. Herr Professor would go to a {\em konditorei}\,\footnote{Pastry shop.} himself  to pick up appropriate pastry. At least, that's what I remeber.
Although later on one of my colleagues, Heisenberg's assistant, assured me that it was his duty to go to {\em konditorei} to bring pastry for tea. It makes sence -- he was born in Vienna  and knew a lot about such things. Most probably I remember the year before his arrival to Leipzig.

Heisenberg loved to play table tennis. He was a very good player. We all played in our spare time. Once a Chinese physicist visited us. 
He was able to beat Heisenberg in table tennis. This became a sensation. I heard later that when Heisenberg sailed from the US to Japan
he was
training all the way to avoid another defeat.

Heisenberg did not like pure mathematics and considered it only as a necessary tool. His method was as follows. Reflecting on a problem he at first would guess its solution, and then would select the mathematical method that would give the conjectured solution. This is a good method if your intuition is as powerful as Heisenberg's. For others this approach is somewhat dangerous.

In Leipzig I managed to write my first paper. It was related to the so-called anomalous Hall effect.

When a metallic plate conducting electric current is put in the perpendicular magnetic field, then a voltage appears in the transverse direction. This is due to deviation of the electrons in the current in the magnetic field. However, in some metals the effect has the opposite sign. Now we know that in such metals the current is not due to the motion of electrons but, rather, that of holes. In 1928 it was not clear. Heisenberg told me that Bloch developed the electron theory of matter and I should apply it to the anomalous Hall effect. To my great pleasure it turned out that indeed this could be done, and I solved the problem the problem.

I spent a year in Leipzig. Heisenberg was invited to the United States, so he took sabbatical. Before leaving he advised me to move to Zurich to work with Pauli. With him I finished my PhD. I should say that I am very grateful to the system of sabbaticals and invitations to the US which allowed me to have such a remarkable combination of teachers.

In addition to his scientific world acclaim Pauli was well-known for his not quite polite remarks. One of the sharpest attacks was made by Pauli in conversation with Ernst Stueckelberg.\footnote{Ernst Stueckelberg (1905-1984) was a Swiss mathematician and physicist known for  causal $S$ matrix theory and the renormalization group. His idiosyncratic style and publication in minor journals led to his work being essentially  unrecognized.}
Addressing Pauli Stueckelberg said: ``Please, do not speak so fast. I cannot think as fast as you." Pauli replied: ``That's OK that you think slowly. I object when you publish faster than you can think."

Somebody showed to Pauli a work of a young theorist being well aware that the work was not too good but still willing to hear Pauli's opinion. Pauli read the paper and said, with sadness: ``It is not even wrong."

\begin{figure}[t]
\epsfxsize=7.5cm
\centerline{\epsfbox{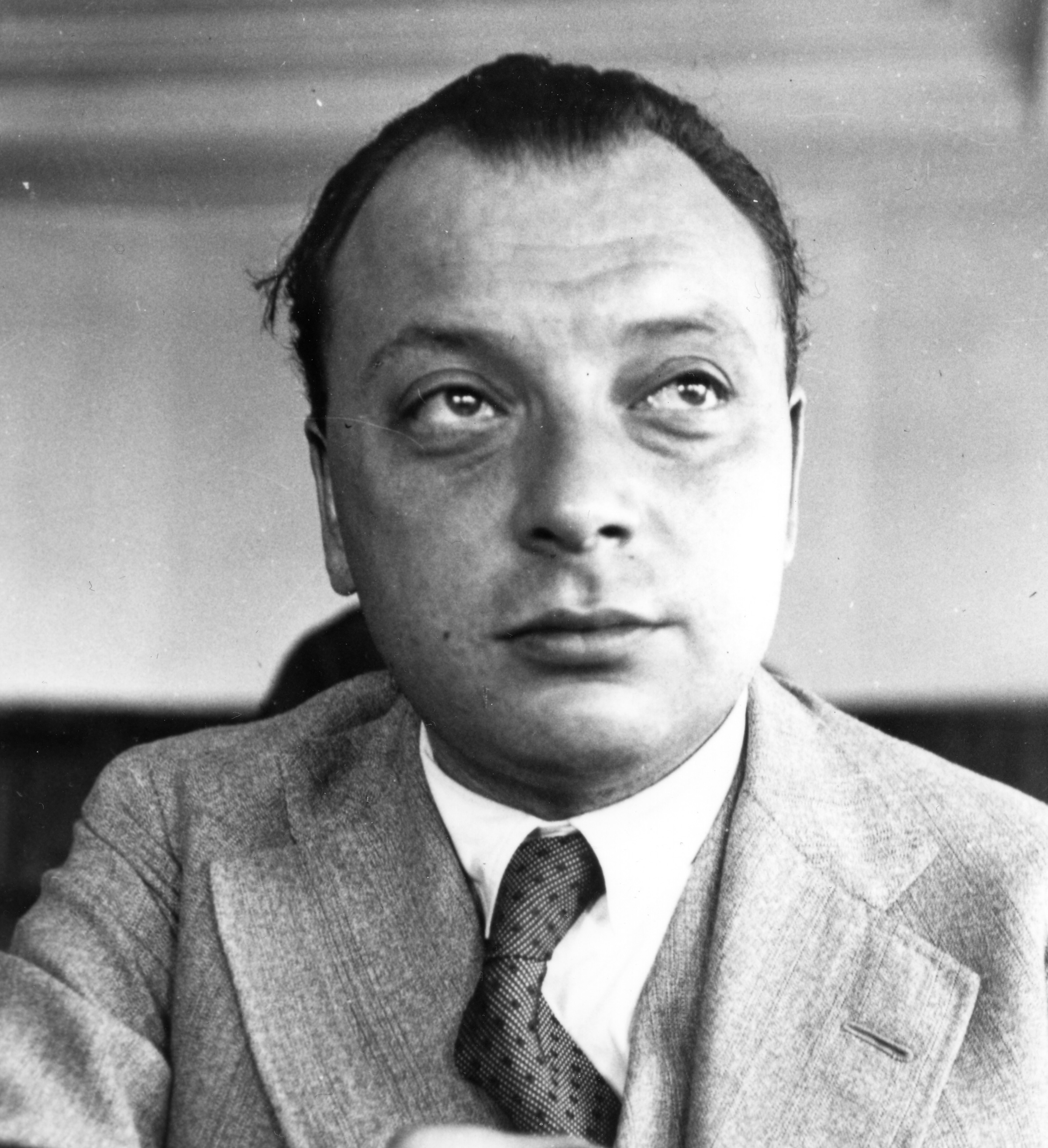}}
\caption{\small
Wolfgang Pauli. Photograph by Francis Simon, courtesy of AIP Emilio Segr\`e Visual Archives, Simon Collection.}
\label{figuk2}
\end{figure}

Wolfgang Pauli. Photograph by Francis Simon, courtesy of AIP Emilio Segrè Visual Archives, Simon Collection

Pauli had a habit. In the evening he would go  to the cinema, to a concert or something like that. Then he would return home by 11 pm and immediately to his desk. Pauli would then work for rather long. Therefore he used to get up late. Once he was invited to a meeting at 9 am. ``Oh, I have to decline, -- said Pauli, -- I can't stay that long without sleep."

One day Pauli was in another city and asked a local physicist how he could find a movie theater. The physicists explained and next day asked Pauli whether he managed to find it. Pauli replied: ``You express yourself quite clearly unless you speak about physics."

I stayed in Zurich working with Pauli for three years. I have heard such things from him more than once. I got used to -- it was not that hard. Nobody took offense at Pauli for a long time, probably because he was also critical of himself, of his own ideas. Once he explained me why he does that. ``To my mind, -- said Pauli, -- there are such sensitive people with whom you can live only by stepping on their `sore corn' a sufficiently large number of times."
I do not think, though, that that was the true reason.

Landau visited Zurich twice. The first time he came in January 1930. At that time Switzerland and the USSR had no diplomatic relations. Landau was granted permit to stay for two weeks, then it was extended by extra two weeks. People tried to help, but at the end of the month he had to leave. Landau joked: ``Lenin was in Switzerland for a few years, and did not ignite revolution. Apparently, they are afraid that I will manage to do that."

In a year he returned with the Rockefeller grant, and then there were no problems -- he could stay as long as he wished. We worked together. Landau was a very young physicist, but very thorough. When a new paper would arrive to our library which interested him, he did not read it. Skipping this stage he would start calculating himself. If he agreed with what was written in the paper he would consider the work good.

Landau liked to order everything. For instance, he divided physicists into different classes:  the first class consisted of Bohr and Sommerfeld, Einstein was in a superior class of his own. Landau modestly put himself to the second class. He classified not only physicists but other aspects of life too.

?andau hated beards. He used to say that beard is a relic of Victorian times, especially on the young faces. In our group we had a physicist 
who had, instead of the beard, long sideburns. Landau considered sideburns as a bourgeois feature too. He called poor fellow's wife and asked:  ``When will you convince your husband  to shave off his ridiculous sideburns?"

Landau insisted that in the West, in Zurich one sees more men with beards than in Leningrad. We made a bet and counted a number beards on a street. Later when I came to Leningrad we performed the same count and found that in Leningrad beards were more common than in Zurich. I won the bet but Landau thought up an explanation: ``Collectivization is under way, and many peasants left their villages and came to  city."

Landau was convinced that only young theorists could do useful things. Later he changed his mind, it's true. Once we had a conversation and a name of a theorist of whom he never heard popped up. Upon learning that he was 27 years old, Landau said:  ``Soo young and already so unknown!"

Besides Landau other people visited Zurich, including George Gamow. Gamow was already famous. He also had a subtle sense of humor and loved all sorts of jokes. Once we went to the mountains and got to the top of a peak with a rather interesting name. There Gamow pulled out a piece of paper from his pocket -- it was a letter to the journal ``Nature" about some kind of nuclear reaction, which he had not completely finished. Gamow siting on the top  finished his letter. Having written the last lines, he put the name of the peak where they were written in the article, and thanked his companions for the opportunity to work there.

In those times, exactly as now, physicists liked to travel, attend conferences and sessions. Travel expenses were not reimbursed at that time, so I often had to sit all night in the corner of a third-class railway carriage. But when I received an invitation to the Physical Society Congress in Odessa, at least inside the country I traveled with great comfort, as a guest. I was invited there by Yakov Ilyich Frenkel, who had read my works.

One of the favorite travel destinations was Copenhagen, where Niels Bohr worked. He was a wonderful person. Bohr hated to offend people, but at the same time he could not allow something contrary to the truth to be said. And from these two qualities a strange combination came out. Once Bohr said: ``I do not say this for criticizing you but what you said is total gibberish." Some other time he also said that clarity and truth are complementary concepts, and indeed, in his works, he rather approached the boundaries of verity.

Bohr's writing process was quite complicated. It began with Bohr dictating, and one of the guests had to write everything down. Then there was editing, the expressions being changed in such a way that everything written was absolutely correct, no ambiguous statements. Many changes folowed, the pages were rewritten, then retyped on a typewriter, then corrected again, etc. Finally, everything was sent to the journal of the Danish Academy, where Bohr's works were, of course, greatly appreciated. After arrival of galley proofs, it was the time for meticulous work on proofreading. The number of consecutive iterations sometimes reached 16. 

Obviously, this was not only Bohr's attitude to words. Once he came to see a new building that was being built for the institute. The contractor, who knew him well, said: ``Professor Bohr, do you see this wall? If you want to move it again, then, please, decide right away, since after three hours the concrete will harden."

As befits a professor, Bohr was rather distracted. I remember that during conversation he smoked cigars all the time. He smokes -- and suddenly asks if there are matches. They give him matches. He tries to light a cigar without interrupting the conversation, which is quite difficult. Then he puts the matches in his pocket, and five minutes later asks the same question again, and everything started all over again. For a long time I kept a smoked piece of chalk: he somehow held a cigar and chalk in one hand.

At this time, troubles began in Germany, and in Copenhagen we discussed not only  physics, but also how to find work for scientists from Germany and Austria. It was not an easy time for scientists in general. Europe was hit hard by economic crisis, universities did not expand, and jobs were vacated only when someone resigned or died. A doctoral dissertation did not guarantee a place for scientific research at all. I had a Rockefeller scholarship for a year, and when I left Zurich I could spend half the time in Rome and half in Cambridge.  Hans Bethe who had spent the winter in Cambridge and the summer in Rome, had already done this before me. I did the opposite and still think I found the best solution.

In Rome, I happened to meet Enrico Fermi, who was also a wonderful physicist. When asked about a problem, he almost always took out a book from the shelf, in which he had already solved this problem. Basically, they were simple -- Fermi did not like complex problems. But then the question arises what to call simple problems? Didn't they become simple after Fermi solved them?

\begin{figure}[t]
\epsfxsize=11cm
\centerline{\epsfbox{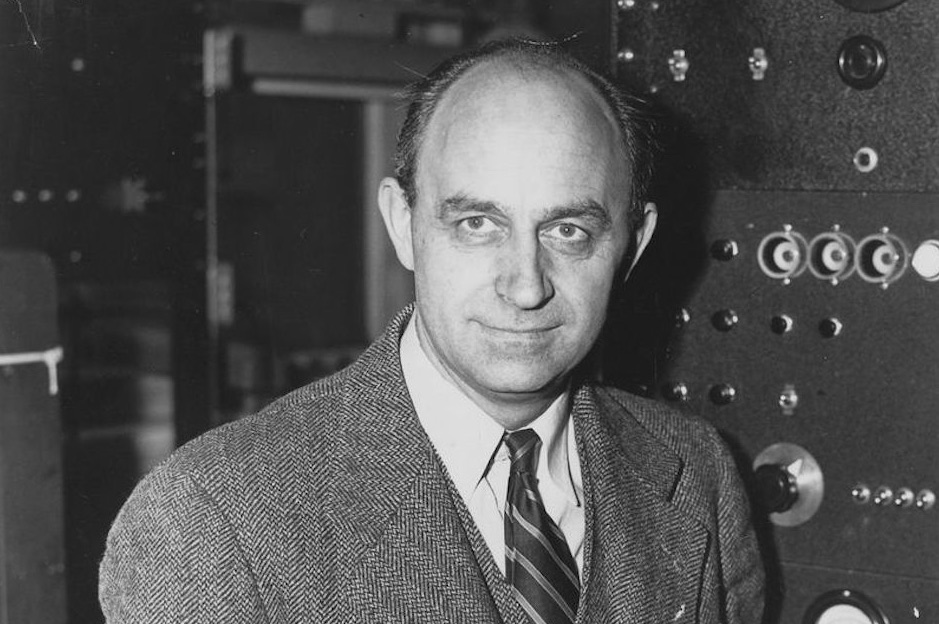}}
\caption{\small
Enrico Fermi.}
\label{figuk3}
\end{figure}

The greatest impression of Fermi came to me later, already in Los Alamos, during the test of the atomic bomb. Everyone, of course, wanted to know what the bomb's power was. There were a lot of tools to determine it, but it required  some time. And Fermi prepared small pieces of paper, and when the blast wave came to us (we were about 15 km from the explosion site), he released these pieces in the air. From the distance that they scattered, he was able to fairly quickly determine the power of the explosion. I do not know what struck me the most then: the idea of the method or the fact that Fermi so precisely determined the moment when it was necessary to release the pieces of paper. I am sure that if I were in his shoe, I would either have released the pieces too early, or I would have forgotten to release them altogether.

After Rome, as I have already mentioned, my wife and I went to Cambridge, where the most interesting contact was with Paul Dirac. Dirac was very polite and treated us with exceptional hospitality. We did not have a car, and he, knowing this, drove us in his, of which he was very proud. They joked that Dirac the driver had a peculiarity: the speed of his car took on only two values -- zero and maximum.

\begin{figure}[t]
\epsfxsize=10cm
\centerline{\epsfbox{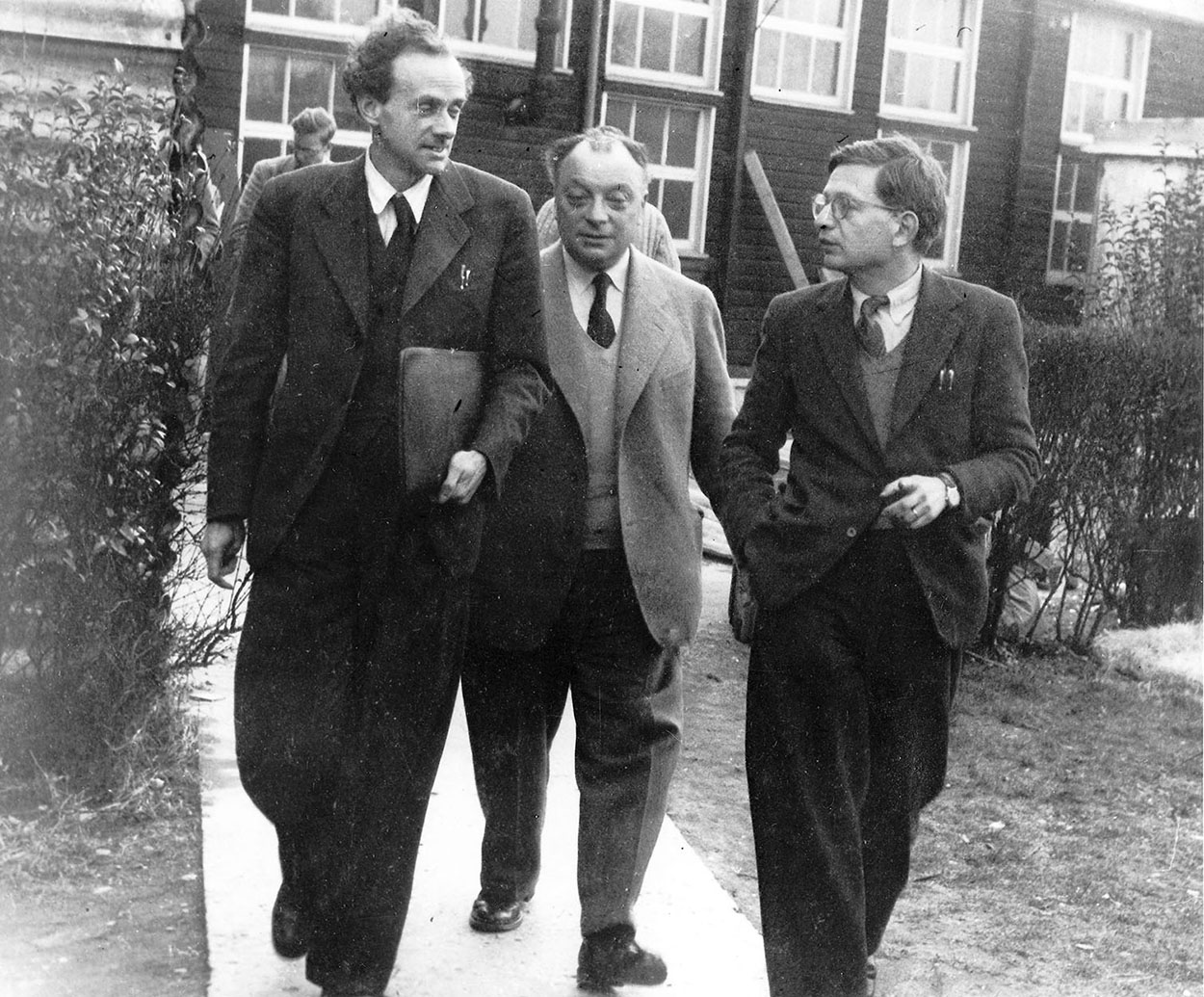}}
\caption{\small
Paul Dirac, Wolfgand Pauli, and Rudolf Peierls, 1953.}
\label{figuk4}
\end{figure}

Dirac has always surprised with his strange reactions. However, if you think about it properly, you'd realize that his words or actions followed absolutely logically from the previous steps. Here's one example. Once a historian of science came to Cambridge and wanted to meet Dirac. They brought him to the College. Dirac was having dinner, and there was a certain silence that had to be somehow defused. The historian started talking about the weather, noting that it was windy outside. Dirac said nothing, then got up, went to the door, opened it and listened. Only after being convinced of the truth of what was said, he expressed his agreement with a monosyllabic ``yes."

Concluding my talk, I would like to remind you that I did not speak as a historian of science, accurately weighing words and placing the right accents. These were the impressions of a witness to the glorious period of the creation of one of the greatest physical theories - quantum mechanics and the memories of its creators, with whom I was lucky to meet and work.

\end{document}